\documentclass{aa}  

\usepackage{graphicx}
\usepackage{txfonts}
\usepackage{natbib,twoopt}
\bibpunct{(}{)}{;}{a}{}{,}
\usepackage{graphicx}
\usepackage{txfonts}
\usepackage{hyperref}
\hypersetup{
    colorlinks=true,
    linkcolor=blue,
    citecolor = blue,
    filecolor=magenta,      
    urlcolor=blue}

\graphicspath{{./}{figures/}}

\begin{document} 

\authorrunning{Billi et al.}
\titlerunning{Fast Rotating Blue Straggler Stars in the globular cluster NGC 1851}

\title{Fast Rotating Blue Straggler Stars in the globular cluster NGC 1851}

\author{A. Billi\inst{1}\fnmsep\inst{2},
    L. Monaco\inst{3}\fnmsep\inst{4},
    F. R. Ferraro\inst{1}\fnmsep\inst{2},
    A. Mucciarelli\inst{1}\fnmsep\inst{2},
    B. Lanzoni\inst{1}\fnmsep\inst{2},
    M. Cadelano\inst{1}\fnmsep\inst{2},
    and I. Trangolao\inst{5}}

\institute{Dipartimento di Fisica e Astronomia, Università degli Studi di Bologna, Via Gobetti 93/2, I-40129 Bologna, Italy\\
\email{alex.billi2@unibo.it}
\and
INAF, Osservatorio di Astrofisica e Scienza dello Spazio di Bologna, Via Gobetti 93/3, I-40129 Bologna, Italy
\and
Universidad Andres Bello, Facultad de Ciencias Exactas, Departamento de F{\'i}sica y Astronom{\'i}a - Instituto de Astrof{\'i}sica, Autopista
Concepc{\'i}on-Talcahuano 7100, Talcahuano, Chile
\and
INAF-OATs, Via G.B.Tiepolo 11, Trieste, I 34143, Italy
\and
Departamento de F{\'i}sica, Universidad de Santiago de Chile, Av. Víctor Jara 3493, Santiago, Chile
}
 
\abstract{In this work we study the rotational velocities of a sample of blue straggler stars (BSSs) and reference stars belonging to the Galactic globular cluster NGC 1851, using high-resolution spectra acquired with FLAMES-GIRAFFE at the ESO/VLT. After field decontamination based on radial velocities and proper motions, the final sample of member stars is composed of 15 BSSs and 45 reference stars populating the red giant and horizontal branches of the cluster. In agreement with previous findings, the totality of reference stars has negligible rotation (lower than 15 km s$^{-1}$). In contrast, we find high values of rotational velocity (up to $\sim$ 150 km s$^{-1}$) for a sub-sample of BSSs. By defining the threshold for fast rotating BSSs at 40 km s$^{-1}$, we found 4 fast-rotating BSSs out of 15, corresponding to a percentage of 27 $\pm$ 14 \%. This results delineates a monotonically decreasing trend (instead of a step function) between the percentage of fast spinning BSSs and the central concentration and density of the host cluster, supporting a scenario where recent BSS formation preferentially occurs in low-density environments from the evolution of binary systems.}

\maketitle

\section{Introduction}
\label{sec:intro}
Blue Straggler Stars (BSSs) are an "exotic" population present in globular clusters (GC; \citealt{Sandage_53}; \citealt{Ferraro+97,Ferraro+99}) and other stellar environments (\citealt{Mathieu_Geller_09}; \citealt{Momany+07}; \citealt{Mapelli+09}; \citealt{Preston_Sneden_00}; \citealt{Clarkson+11}). In the color-magnitude diagram (CMD) they lie on an extension of the Main Sequence (MS) and they are bluer (hotter) and brighter than the MS turnoff. Hence, they are more massive than MS stars (e.g., \citealt{Shara+97}; \citealt{Geller_Mathieu_11}; \citealt{Fiorentino+14}). 
\\
Because of their large mass, BSSs suffer the effect of dynamical friction, which make them progressively migrate in the central regions of the cluster. In this respect, \citet{Ferraro+12} presented the concept of "dynamical clock", a method that uses the BSS radial distribution to derive the dynamical age of star clusters. A parameter that efficiently traces the dynamical evolution of the host cluster is the $A^+_{rh}$ parameter (\citealt{Alessandrini+16}; \citealt{Lanzoni+16}), which is defined as the area between the cumulative radial distribution of BSSs and that of a lighter stellar population of the cluster, such as MS, red giant branch (RGB), or horizontal branch (HB) stars. Therefore this parameter describes the BSS central segregation with respect to that of the lighter population. Since only one burst of star formation occurred in globular (and open)  clusters, mass-enhancement processes must be at the origin of these peculiar objects. In this respect, three formation scenarios have been proposed so far: mass transfer from a companion star to the proto-accreting BSS in binary systems (MT-BSS; \citealt{McCrea_64}), direct collisions between two or more stars (COL-BSS; \citealt{Hills_Day_76}; \citealt{Sills+05}), and mergers or collisions triggered by dynamical interactions or stellar evolution in triple star systems (\citealt{Andronov+06}; \citealt{Perets_Fabrycky_09}). 
\\ 
The MT channel is favored in low-density environments, possibly because binary systems can more easily survive in low-density conditions and in fact \citet{Mathieu_Geller_09} found that $76\pm 19$\% (a fraction that is compatible with 100\%) of BSSs in the old open cluster NGC 188 are in binary systems. In addition, \citet{Sollima+08} found an intriguing correlation between the BSS specific frequency and the binary fraction in the core of thirteen low-density Galactic GCs. Moreover, recently \citet{Ferraro+25} found evidence that the vast majority of BSSs detected in Galactic GCs have a binary-related origin (see also \citealt{Knigge+09}), thus indicating the MT process as the most efficient formation channel also in GCs.
\\
Distinguishing between MT-BSSs and COL-BSSs from observations is a hard task. Photometrically, a far ultraviolet excess has been detected in a few binary BSSs in NGC 188 (\citealt{Gosnell+14,Gosnell+15,Subramaniam+16}) and other open and globular clusters \citep[e.g.][and references therein]{Sahu+19, Dattatrey+23, Reggiani+25}. This has been interpreted as the signature of hot white dwarf companions, remnants of the MT activity that generated these BSSs. Moreover, the BSS population in a few post core collapse GCs has been observed to be split in two well-separated and parallel sequences (\citealt{Ferraro+09}; \citealt{Dalessandro+13}; \citealt{Beccari+19}; \citealt{Cadelano+22}; see also \citealp{Simunovic+14} and \citealp{Raso+20}). The bluest sequence is thought to be mainly populated by COL-BSSs formed in a sudden enhancement of the collision activity during the core collapse phase, while the reddest one is interpreted as predominantly due to MT-BSSs \citep[see]{Ferraro+09, Xin+15, Portegies_Zwart_19}. However, the presence of a double BSS sequence has been claimed also in some open clusters (see, e.g, \citealt{Rao+23}, \citealp{Wang+25}, and references therein, but see also \citealt{Dalessandro+19}), which challenges the previous interpretation since the low-density environment of open clusters disfavors the formation of COL-BSSs. One intriguing suggestion \citep{Rao+23} is that one sequence is populated by slowly rotating stars and the other one by fast rotating stars. Further investigations are needed to confirm the hypothesis. Another possibility suggested by the authors is the presence of multiple populations.
\\
An alternative way for distinguishing MT-BSSs from those formed via collision is through the spectroscopic study of the surface chemical composition. In fact, MT-BSSs are predicted to show surface depletion in carbon and oxygen due to the accretion of material that was processed deep into the interior of the companion star  (\citealt{Sarna_DeGreve_96}), while no chemical anomalies are expected in the COL scenario (\citealt{Lombardi+95}). The chemical signature of MT formation has been observed in sub-samples of BSSs belonging to the GCs 47 Tucanae (\citealt{Ferraro+06}) and M30 (\citealt{Lovisi+13a}). Interestingly, both the chemical (CO-depletion) and photometric (UV excess) signatures have been detected for the first time in the same BSS in 47 Tucanae \citep{Reggiani+25}. Moreover, recent studies in open clusters detected a barium enhancement in a sub-sample of BSSs (\citealt{Milliman+15}; \citealt{Nine+24}). This feature has been interpreted as the signature of MT between an asymptotic giant branch companion that polluted the surface of the BSS with elements produces during the thermal pulses. 

\begin{figure}[h!]
\centering
\includegraphics[width=\columnwidth]{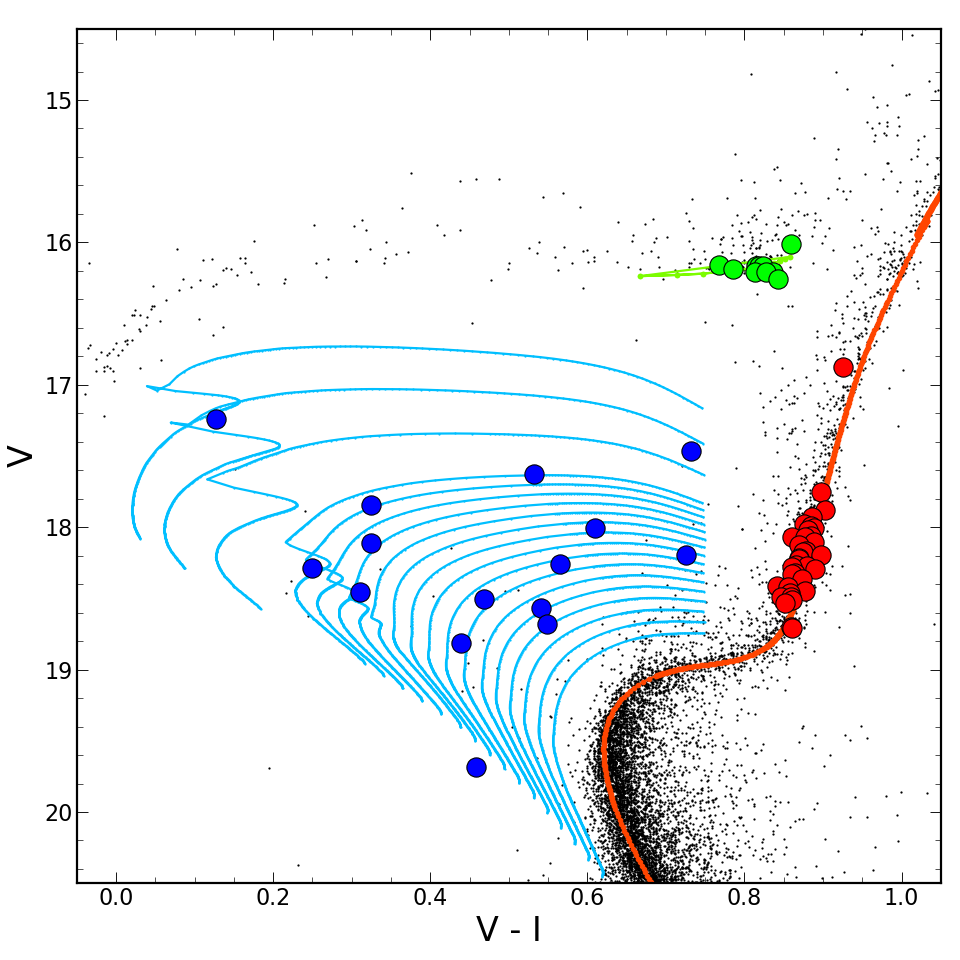}
\caption{CMD of NGC 1851 (black dots) with the surveyed BSSs, RGB and HB stars marked, respectively, as blue, red, and green circles. A set of BASTI evolutionary tracks (\citealt{Pietrinferni+21}) with masses ranging from 0.9 to 1.5 $M_\odot$ are overplotted as cyan lines. A BASTI isochrone of 11 Gyr is overplotted as orange line, and a BASTI HB model is overplotted as green line.}
\label{fig:cmd}
\end{figure}

\begin{figure}[h!]
\centering
\includegraphics[width=\columnwidth]{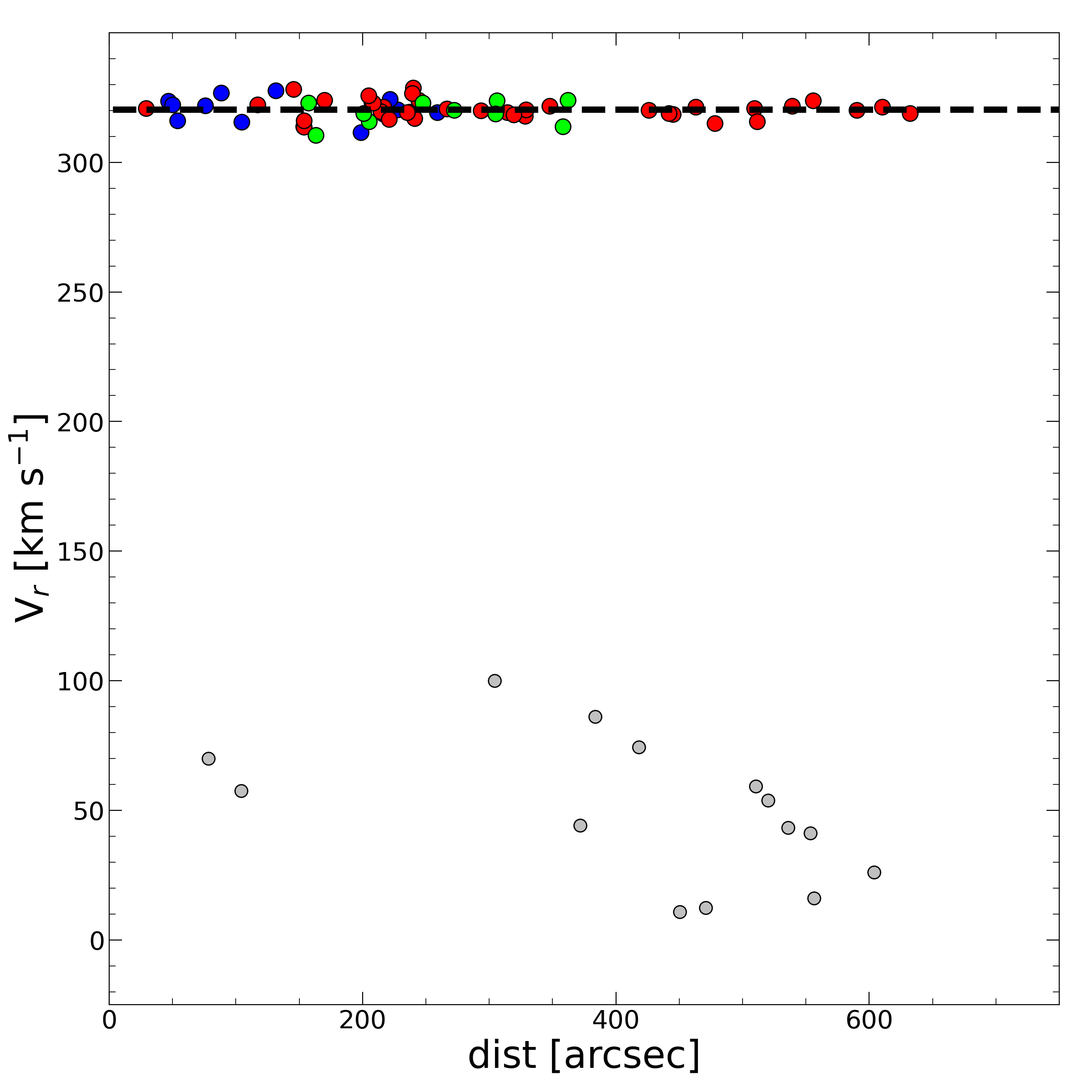}
\caption{Radial velocity of the observed stars as a function of their distance from the cluster center. Member BSSs, RGB and HB stars are marked as blue, red and green circles, respectively. The gray circles correspond to Galactic field interlopers. The black dashed line marks the average radial velocity of the member star sample.}
\label{fig:rv}
\end{figure}

\noindent The study of rotational velocities can also add important information on BSS origin, and could shed light on the link between the BSS evolution and the environmental characteristics. In particular, high rotational velocities are expected at birth for both MT-BSSs (\citealt{Sarna_DeGreve_96}; \citealt{deMink+13}) and COL-BSSs (\citealt{Benz_Hills_87}; \citealt{Sills+02}). For the MT channel, the reason for high velocity is the transfer of angular momentum from the companion star to the accreting proto-BSS, whereas in the COL scenario the reason is the conservation of angular momentum. Braking mechanisms (e.g., magnetic braking and disk locking) should then slow down the stars, with timescales and efficiencies that are not completely understood yet (\citealt{Leonard_Livio_95}; \citealt{Sills+05}). Luckily, new constraints are emerging from observations in open and globular clusters. A recent result in open clusters compares the rotation rates and the ages of MT-products to model the evolution of angular momentum with time, suggesting a timescale shorter than 1 Gyr for substantial braking (\citealt{Leiner+18}). A similar value (1-2 Gyr) has been inferred \citep{Ferraro+23b} for COL-BSSs using the time since core collapse (estimated from the comparison between collisional models and the observed blue BSS sequence in M30; see \citealp{Ferraro+09}) as an estimate of the formation epoch of these stars. Hence, although the rotational velocity cannot univocally indicate the formation scenario,  
it can be used as indicator of BSS age.
\\
In this context, several years ago our group started an extensive high-resolution spectroscopic campaign to study the rotational velocities of these exotic stars in a sample of Galactic GCs with different values of the structural parameters.
\\
So far, we published the results about BSS rotational velocities in eight systems, namely 47 Tucanae (\citealt{Ferraro+06}), M4 (\citealt{Lovisi+10}), NGC 6397 (\citealt{Lovisi+12}), M30 (\citealt{Lovisi+13a}), NGC 6752 (\citealt{Lovisi+13b}), $\omega$ Centauri\footnote{Although $\omega$ Centauri is likely the remnant of a nuclear star cluster of an accreted dwarf galaxy (\citealp{Bekki_Freeman_03}), the properties of its current BSS population should not depend on its true origin and can be safely compared to those of genuine GCs.} (\citealt{Mucciarelli+14}), NGC 3201 (\citealt{Billi+23}), and M55 (\citealt{Billi+24}). Using this dataset, which includes about 300 BSSs, \citet{Ferraro+23a} found a strong relation between the fraction of fast rotating BSSs (FR-BSSs) and the structural parameters (e.g., King concentration, central density) of the host cluster, with FR-BSSs being defined as those with v sin(i) $\geq$ 40 km s$^{-1}$, i being the inclination angle in the plane of the sky\footnote{As discussed in \citet{Ferraro+23a}, the choice of the 40 km s$^{-1}$ threshold has been done considering the observed rotational velocities distributions, since it has been noted that some clusters show a clear drop in the fraction of FR-BSSs in correspondence with this value. It has been also demonstrated that choosing a slightly different threshold (such as 30 or 50 km s$^{-1}$) has no impact on the results.}. For low-density and low-concentration clusters the fraction of FR-BSSs is much higher than the one observed in dense and more concentrated systems. This suggests an influence of the environment on the formation and evolution of BSSs (see also \citealt{Ferraro+25}).
\\
In this paper we present new results on the rotational velocities of BSSs and of a reference sample composed of RGB and HB stars in the GC NGC 1851. This cluster has an intermediate value of the concentration ($c = 1.86$) and a high value of central density (log $\rho_0$ = 5.09 in units of L$_{\odot}$ pc$^{-3}$). It is the first one in our sample with intermediate King concentration where the BSS rotational velocities have been measured, thus filling the gap between high- and low-concentration systems. 
The paper is organized in the following sections. Sect. \ref{sec:obs} describes the observations and data reduction. The determination of atmospheric parameters is described in Sect.\ref{sec:atmospheric_parameters}. Sect. \ref{sec:rv_membership} discusses the membership of stars based on radial velocities and proper motions. In Sects. \ref{sec:rotational_velocities} and \ref{sec:discussion}, we present the results obtained for the rotational velocity of BSSs and reference stars and the conclusions of the work, respectively.

\begin{table*}
\caption{Properties of the analyzed BSSs.}
 \label{table:1}
    \centering 
    \begin{tabular}{c c c c c c c c c}
        \hline\hline 
        ID & ID Gaia & R.A. & Dec. & V & (V$-$I) & V$_r$ & v sin(i) \\
         &  & [deg] & [deg] & & & [km s$^{-1}$] & [km s$^{-1}$] \\
        \hline 
        23317 & 4819197883024929280 & 78.5151667 & -40.0114722 & 18.50 & 0.47 & 327.7 $\pm$ 1.0 & 16 $\pm$ 2 \\ 
        42825 & 4819291547673927808 & 78.5921250 & -39.9938333 & 18.57 & 0.54 & 319.3 $\pm$ 1.0 & < 15 \\  
        44048 & 4819279693564194432 & 78.6020000 & -40.0180556 & 17.63 & 0.54 & 320.3 $\pm$ 3.9 & 69 $\pm$ 4 \\
        44561 & 4819279620550033024	& 78.6068333 & -40.0337500 & 18.68 & 0.55 & 324.4 $\pm$ 1.4 & < 15 \\
        34141 & 4819197608151426304 & 78.5500417 & -40.0338611 & 18.00 & 0.61 & 321.9 $\pm$ 0.6 & < 15 \\
        28789 & 4819197608151755648 & 78.5322083 & -40.0339444 & 17.50 & 0.74 & 323.7 $\pm$ 0.5 & 15 $\pm$ 2 \\
        43811 & 4819279590484996096 & 78.6000000 & -40.0508333 & 18.20 & 0.73 & 311.5 $\pm$ 0.7 & < 15 \\
        32911 & 4819197539428458112	& 78.5457917 & -40.0433333 & 17.84 & 0.32 & 322.2 $\pm$ 0.8 & < 15 \\
        24367 & 4819197470712614272 & 78.5185833 & -40.0595556 & 17.24 & 0.13 & 316.1 $\pm$ 1.4 & 65 $\pm$ 5 \\
        18643 & 4819197058394790528 & 78.4984583 & -40.0913333 & 18.11 & 0.33 & 287.8 $\pm$ 8.8 & 147 $\pm$ 16 \\
        14237 & 4819196955314217600 & 78.4785417 & -40.0931389 & 19.69 & 0.46 & 318.8 $\pm$ 1.6 & < 15 \\
        10013 & 4819197951749529600 & 78.4477917 & -40.0700278 & 18.82 & 0.44 & 319.4 $\pm$ 1.6 & 33 $\pm$ 7 \\
        20558 & 4819197779951454976 & 78.5057083 & -40.0290278 & 18.26 & 0.56 & 326.8 $\pm$ 1.0 & < 15 \\
        17621 & 4819197745587678592 & 78.4943333 & -40.0335278 & 18.46 & 0.31 & 315.6 $\pm$ 2.8 & 68 $\pm$ 3 \\
        15834 & 4819198501500313216 & 78.4865417 & -40.0183333 & 18.29 & 0.25 & 313.9 $\pm$ 1.5 & 19 $\pm$ 3 \\
        \hline 
    \end{tabular}
    \tablefoot{Identification number of the Stetson and Gaia catalogs, J2000 coordinates, V-band magnitude and (V$-$I) color (from \citealt{Stetson+19}), radial velocity and rotational velocity of the analyzed BSSs.}
\end{table*}

\section{Observations}
\label{sec:obs}
This work is based on stellar spectra obtained with the high resolution multi-object spectrograph FLAMES-GIRAFFE (\citealt{Pasquini+02}) mounted on the Very Large Telescope at European Southern Observatory (ESO) Paranal Observatory, under programmes 090.D-0487(A) and 092.D-0477(A) (PI: Simunovic). The observations have been performed in different nights between November 2012 and October 2013, using the HR9A setup, covering the wavelength interval $\Delta \lambda$ = 5095 - 5404 $\AA$, which samples the Mgb triplet lines ($\lambda_1$ = 5167.4 $\AA$, $\lambda_2$ = 5172.7 $\AA$, $\lambda_3$ = 5183.6 $\AA$). These three lines are very sensitive to v sin(i),  as demonstrated in recent studies on BSS rotational velocities (see \citealt{Mucciarelli+14}; \citealt{Billi+23}). A total of 15 exposures are available in the ESO archive, and the acquired spectra belong to stars evolving along different evolutionary stages: BSSs, RGB and HB stars (see Figure \ref{fig:cmd}, where the adopted photometric catalog is from \citealt{Stetson+19}). We took the reduced spectra from the online ESO archive\footnote{\url{https://archive.eso.org/cms.html}}. Then, we computed a master-sky as the median of the sky spectra in the available exposures, and we subtracted it from the observed spectra. Finally, we 
corrected for the heliocentric velocity and averaged the individual exposures of the same stars. The resulting signal-to-noise ratio (S/N) ranges between 10 and 30 for the sampled BSSs and RGB stars, while it is between 30 and 50 for the HB stars.

\begin{figure} [h!]
\centering
\includegraphics[width=\columnwidth]{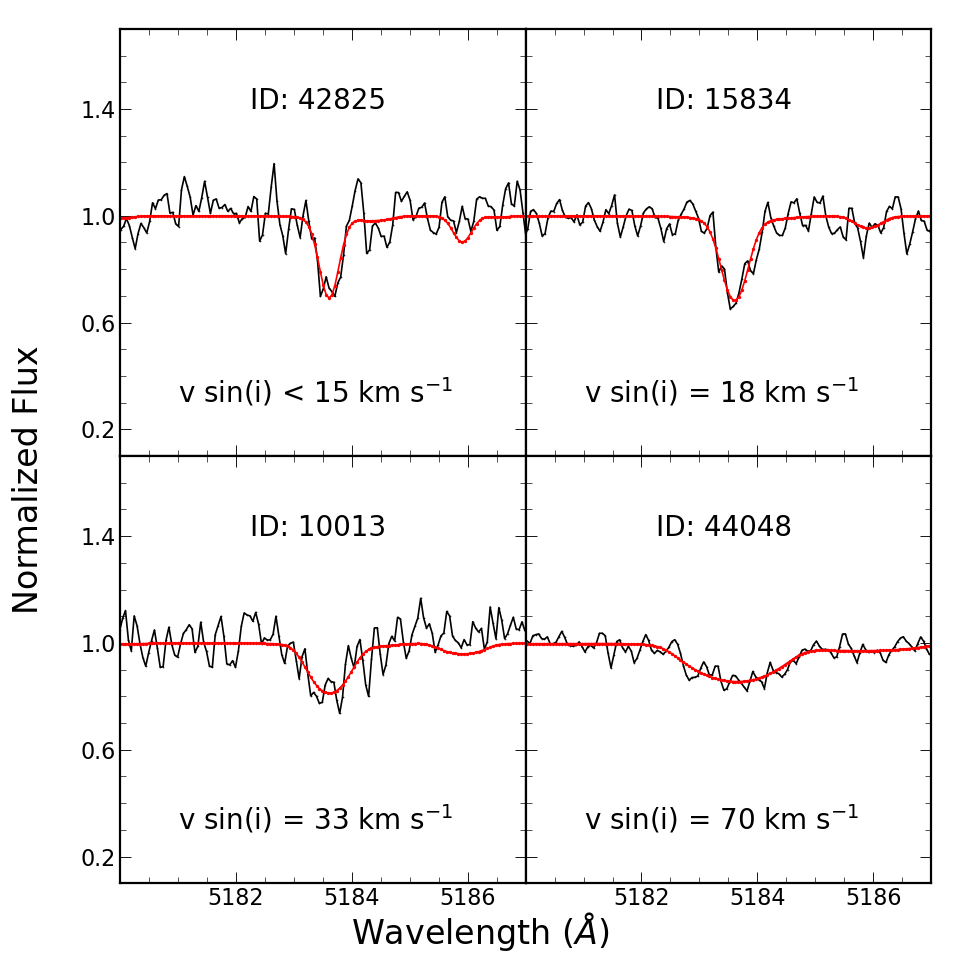}
\caption{Comparison between the observed spectrum (black line) and the best-fit synthetic spectrum (red line) of four BSSs with different rotational velocities (see labels).}
\label{fig:vrot_obs_sint}
\end{figure}

\begin{figure} [h!]
\centering
\includegraphics[width=\columnwidth]{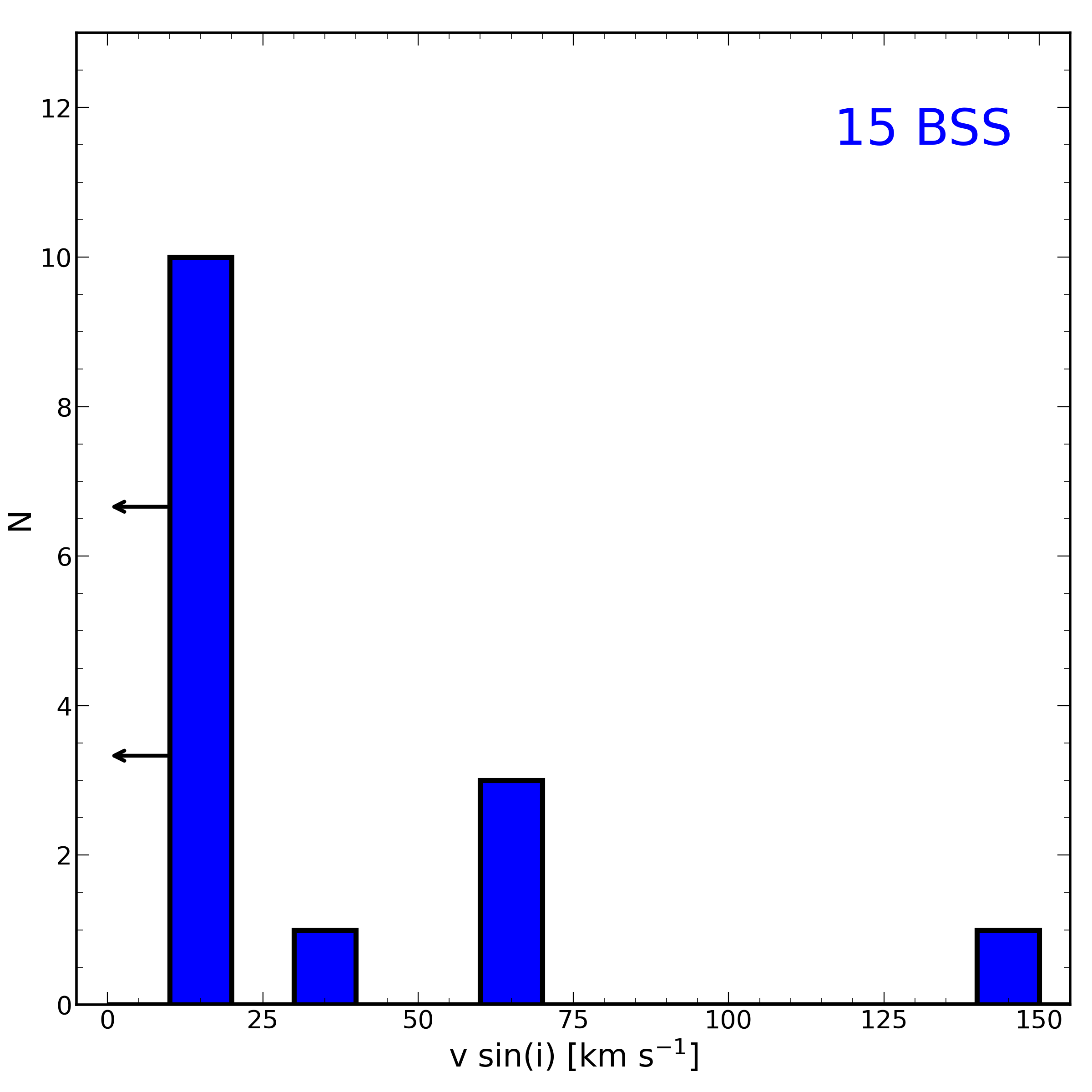}
\caption{Rotational velocity distribution of the 15 BSSs analyzed in this paper. Left arrows indicate the upper limit of 15 km s$^{-1}$ adopted for very slow rotating BSSs.}
\label{fig:vsini_distributions}
\end{figure}

\section{Atmospheric parameters}
\label{sec:atmospheric_parameters}
Atmospheric parameters (such as effective temperature, T$_{\rm eff}$, and surface gravity, $\log g$) are a fundamental ingredients for the computation of appropriate synthetic spectra to be compared with the observed ones, from which radial and rotational velocities are then measured. They have been estimated by comparing the position of the surveyed stars in the CMD with evolutionary tracks, isochrones and an HB model. The models have been retrieved from the online BASTI-IAC database (\citealt{Pietrinferni+21}), adopting an $\alpha$-enhanced mixture and a metallicity [Fe/H] $= -1.18$ dex \citep{Harris_96}. A set of evolutionary tracks with mass values ranging from 0.9 to 1.5 M$_{\odot}$ has been used for BSSs, while the atmospheric parameters of HB and RGB stars have been estimated through the comparison with an HB model including stellar masses of 0.65-0.8 M$_{\odot}$ and an 11 Gyr old (\citealt{Vandenberg+13}) isochrone, respectively. The models have been plotted in the observed CMD (see Figure \ref{fig:cmd}) by adopting a distance modulus $(m-M)_V$ = 15.47 and reddening $E(B-V) = 0.02$ (\citealt{Harris_96}).
\\
To estimate the BSS temperatures and surface gravities we orthogonally projected the observed CMD position of each star onto the closest evolutionary track. For BSSs the resulting temperatures and $\log g$ values range between 5700 and 8600 K, and between 3.4 and 4.2 dex, respectively. For RGB stars we obtained T$_{\rm eff}$ values between 5030 and 5350 K, and $\log g = $[3.3, 3.6] dex. Finally, for HB stars we find T$_{\rm eff}$ = [5300, 5600] K and $\log g =$ [2.4, 2.5] dex. We assumed a microturbolence of 1 km s$^{-1}$ for BSSs and HB stars, whereas for RGB stars we adopted 1.5 km s$^{-1}$. 

\begin{figure}
\centering
\includegraphics[width=\columnwidth]{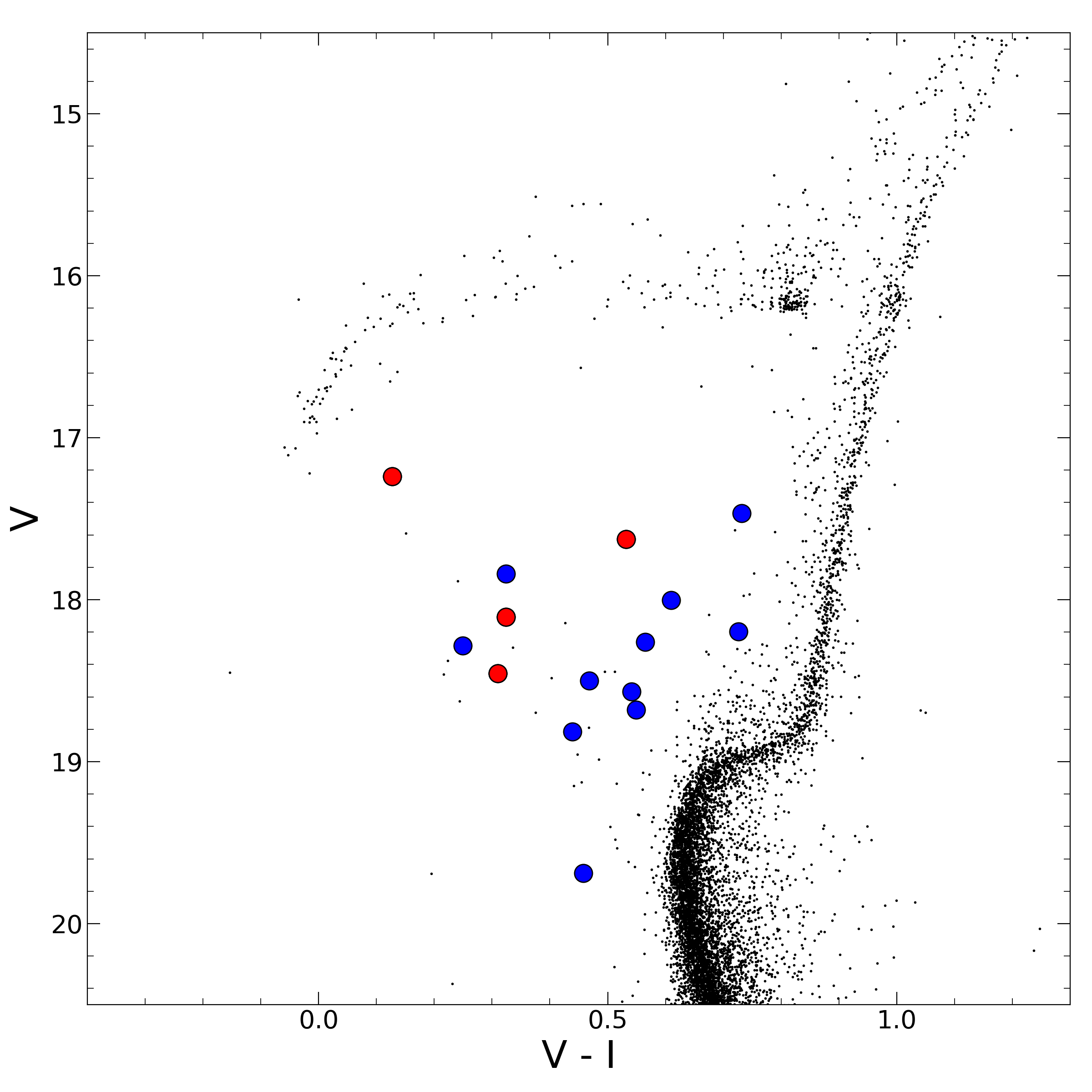}
\caption{Optical CMD of NGC 1851 with the position of fast and slowly rotating BSSs analyzed in this work highlighted with red and blue circles, respectively.}
\label{fig:cmd_fr}
\end{figure}

\section{Radial velocities and cluster membership}
\label{sec:rv_membership}
Radial velocities (V$_r$) have been used to assess the cluster membership of the observed stars. We measured them by cross-correlation between the observed and template spectra, using the IRAF task \textit{fxcor} (\citealt{Tonry_Davis_79}). As templates we used synthetic spectra calculated using the SYNTHE code (\citealt{Sbordone+04}; \citealt{Kurucz_05}) and assuming different atmospheric parameters for the different kinds of stars. For the atomic and molecular transitions we used the last version of the Kurucz/Castelli linelist\footnote{\url{https://wwwuser.oats.inaf.it/castelli/linelists.html}}. The model atmospheres were calculated with the ATLAS9 code (\citealt{Kurucz_93}; \citealt{Sbordone+04}) under the assumptions of local thermodynamic equilibrium and plane-parallel geometry, adopting the opacity distribution functions by \citet{Castelli_Kurucz_03} with no inclusion of the approximate overshooting prescription (\citealt{Castelli+97}).

\begin{figure}
\centering
\includegraphics[width=\columnwidth]{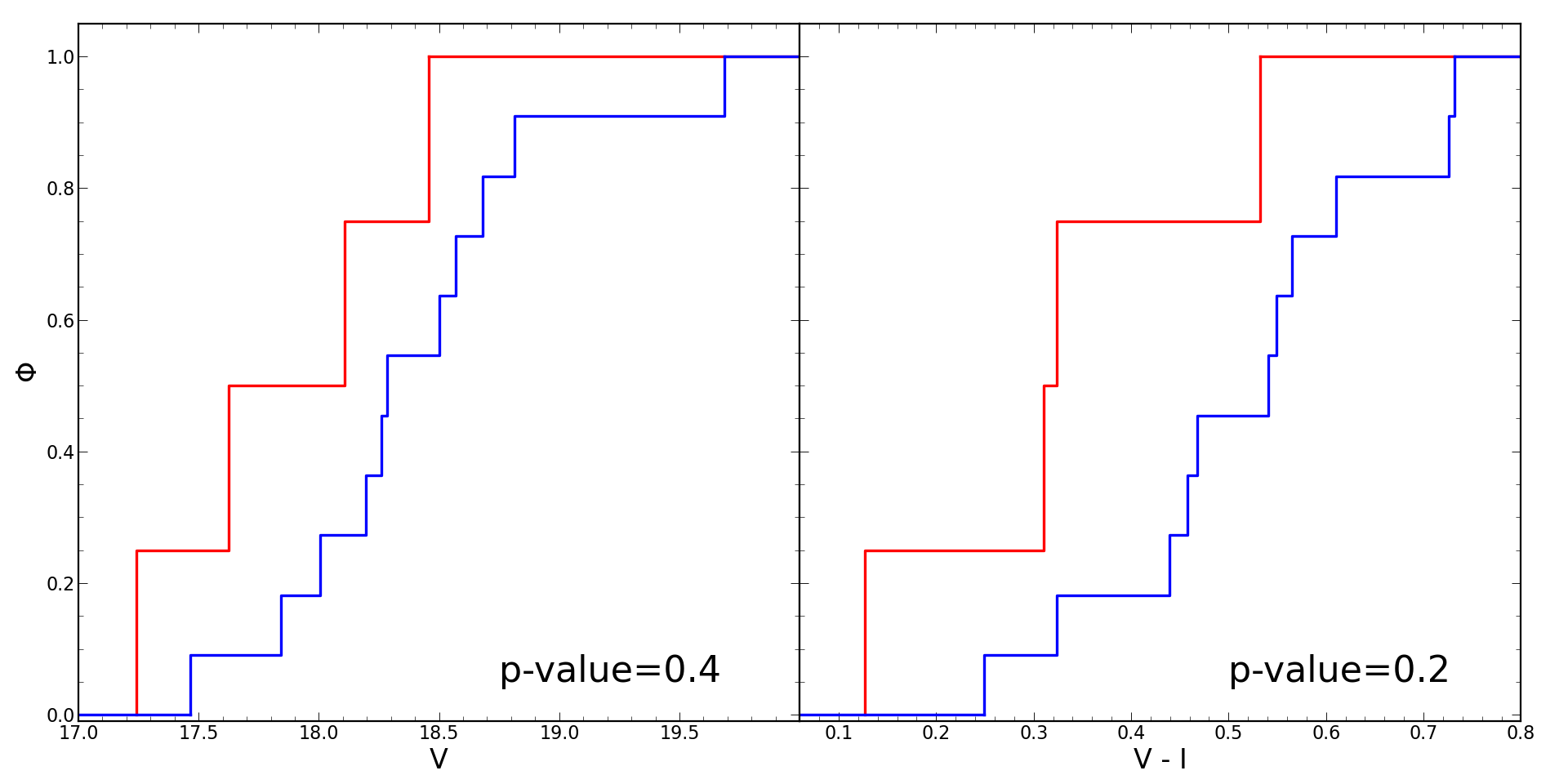}
\caption{Cumulative distributions of the V-band magnitude (left panel) and (V$-$I) color (right panel) for the sample of FR-BSSs (red
lines) and slowly rotating BSSs (blue lines). The p-values of the Kolmogorov-Smirnov test against the null hypothesis that the two samples are extracted from the same parent distribution are marked in each panel.}
\label{fig:cumulativa}
\end{figure}

\noindent Figure \ref{fig:rv} shows the distribution of the measured radial velocities as a function of the distance from the center. As can be seen, 14 objects have radial velocities inconsistent with that of the bulk population and are clearly Galactic field interlopers. Hence, they have been excluded from the following analysis. The remaining stars tightly align around a mean radial velocity of 320.3 $\pm$ 0.5 km s$^{-1}$ (black dashed line in Figure \ref{fig:rv}), a value that is well consistent with the systemic velocity of NGC 1851 quoted in literature: 320.5 $\pm$ 0.6 km s$^{-1}$ \citep{Harris_96} and 321.4 $\pm$ 1.6 km s$^{-1}$ \citep{Baumgardt_Hilker_18}. They all are with 3 $\sigma$ of the distribution, with $\sigma$ = 3.9 km s$^{-1}$. To further confirm their cluster membership, we took into account the membership probability quoted by \citet{Vasiliev_Baumgardt_21} and 
estimated from Gaia DR3 proper motions (\citealt{GaiaCollaboration_16, GaiaCollaboration_23}), finding that they all have membership probability larger than 90\%. For one of the surveyed BSSs we measure V$_r$ = 287.8 km s$^{-1}$. However, this is a very FR-BSS and, because its spectral lines are extremely broadened, the obtained value is not reliable. However, this star (Gaia ID: 4819197058394790528) has membership probability $> 0.9$ in \citet{Vasiliev_Baumgardt_21}, and we therefore kept it in the final sample of bona fide members, which includes a total of 15 BSSs, 35 RGB and 10 HB stars.

\section{Rotational velocities}
\label{sec:rotational_velocities}
Rotational velocities have been measured by comparing the normalized observed spectra with a grid of synthetic spectra calculated with the atmospheric parameters discussed above and with different values of v sin(i). More specifically, we selected the Mgb triplet lines, in particular we used the third line ($\lambda_3 = 5183.6 \ \AA$) that allows the best and cleanest comparison. A $\chi^2$ minimization procedure has been used to identify the best-fit solution. Figure \ref{fig:vrot_obs_sint} shows the comparison between the observed spectrum and the best-fitted synthetic spectrum for the third line of the Mgb triplet, for a sample of four BSSs with different values of rotational velocities. Clearly, the broader is the observed line, the higher is the stellar rotational velocity. To estimate the uncertainties on v sin(i) we performed a Monte Carlo simulation. For each star we calculated a synthetic spectrum with the proper atmospheric parameters and the best-fit value of v sin(i). Then, we randomly added Poissonian noise with the same S/N ratio of the observed spectrum and repeated the analysis described above, thus deriving a new best-fit value of v sin(i). By repeating the procedure 300 times, we obtained a distribution of v sin(i) values, and finally adopted its standard deviation as the 1$\sigma$ uncertainty on the rotational velocity of each star. 
Table \ref{table:1} lists the values obtained for each member BSS. Due to the moderate resolution and relatively low S/N of the spectra, we are not able to distinguish very slow rotating stars (with v sin(i) < 15 km $s^{-1}$). Therefore we adopt an upper limit for those stars. RGBs and HBs, as expected, show very slow rotation, with v sin(i) < 15 km s$^{-1}$. This is consistent with findings of previous studies (\citealt{Billi+23,Billi+24}; \citealt{Ferraro+23a}).
\\
The distribution of v sin(i) values obtained for the surveyed sample of BSSs is shown in Figure \ref{fig:vsini_distributions}. As can be seen, the rotational velocity distribution of BSSs is different from those of "normal" cluster stars. As said before, all RGB and HB stars rotate more slowly than 15 km s$^{-1}$, whereas some BSSs not only have v sin(i) values above this threshold, but  even show rotational velocities exceeding 100 km s$^{-1}$. 
\\
In particular, in NGC 1851 we find 4 FR-BSSs out of the 15 BSSs analyzed, which corresponds to a total fraction of 27 $\pm$ 14\%. The uncertainty on the fraction of FR-BSSs has been calculated using the propagation of uncertainty of a ratio.

\begin{figure*}
\centering
\includegraphics[width=2\columnwidth]{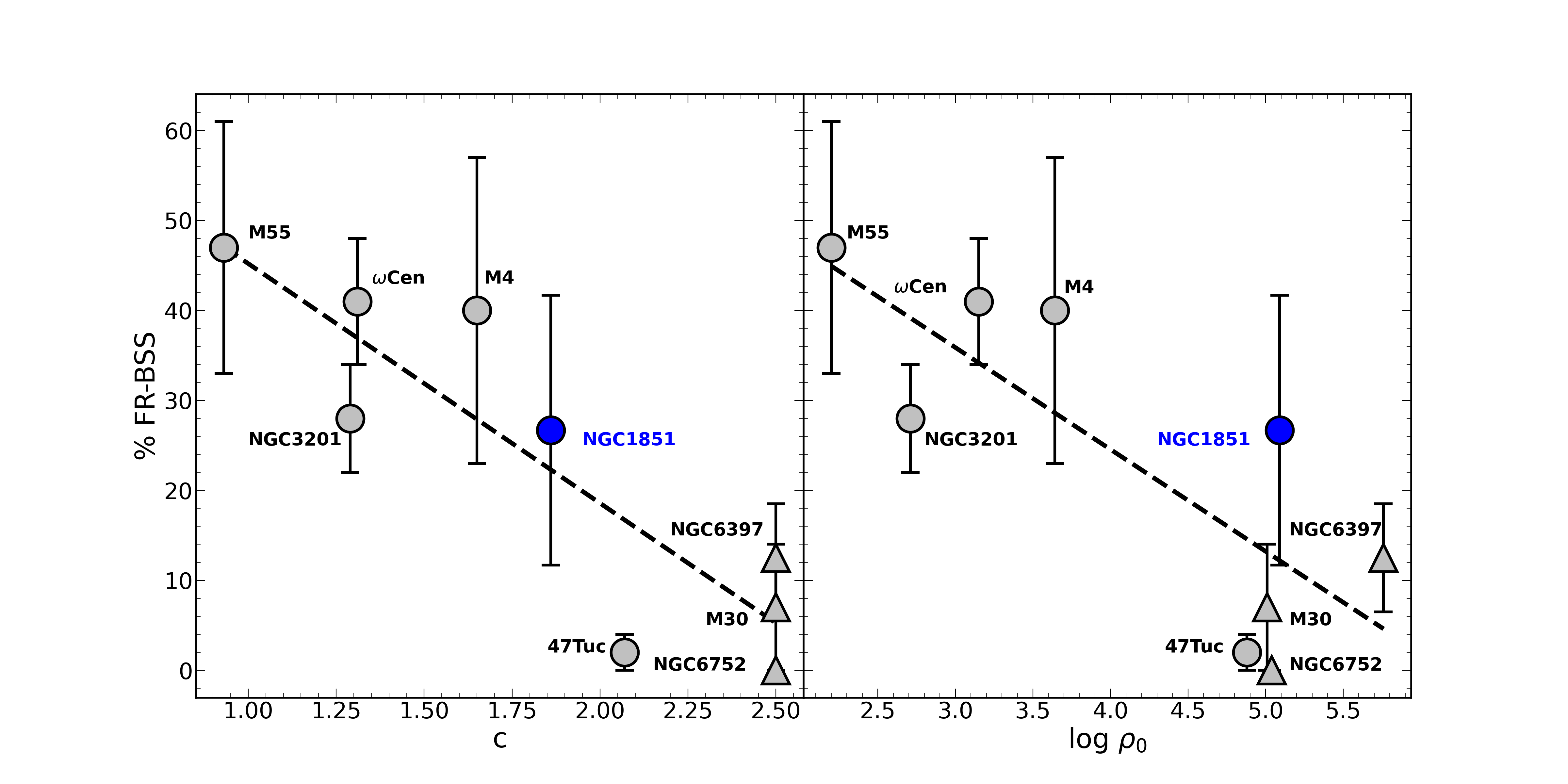}
\caption{Percentage of FR-BSSs as a function of the concentration (left panel) and the central density (right panel) of the parent cluster. The gray symbols correspond to the GCs investigated in \citet{Ferraro+23a}, while the blue circle corresponds to NGC 1851. Triangles represent post core collapse GCs.}
\label{fig:grafico_percentuale}
\end{figure*}

\section{Discussion and conclusions}
\label{sec:discussion}
As part of a spectroscopic survey of BSS rotational velocities in a sample of Galactic GCs spanning a large range in structural parameters, this work presents the results obtained in NGC 1851. Similar to other clusters, we find that BSSs have a peculiar rotational velocity distribution compared to standard cluster stars, enforcing their "exotic" origin. In particular, BSSs with v sin(i) of the order of 100 km s$^{-1}$ or more have been detected in previous works (\citealt{Lovisi+10}; \citealt{Mucciarelli+14}; \citealt{Billi+23,Billi+24}) and one BSS with an extremely high rotational velocity (almost 150 km s$^{-1}$) is also found in NGC 1851. 
\\
Figure \ref{fig:cmd_fr} shows the position in the CMD of slowly and fast rotating BSSs, while in Figure \ref{fig:cumulativa} we plot the cumulative distributions of the two sub-samples as a function of the observed magnitude and color (left and right panels, respectively). Figure \ref{fig:cumulativa} also provides the p-values of the Kolmogorov-Smirnov test against the null hypothesis that the two samples are extracted from the same parent distribution. Unfortunately, the small size of the available samples does not allow us to obtain conclusive answers about the possible difference in the color and magnitude distributions of FR and slow-rotating BSSs.
\\ 
Since a high rotational velocity is interpreted as the evidence of recent BSS formation, with braking mechanisms that had not enough time to intervene and slow down the stars, this piece of evidence might suggest that brighter and bluer BSSs tend to be younger than the others, or that braking mechanisms tends to be less efficient in the former. We looked also for a possible correlation between the fraction of FR and the distance of the BSSs from the center of the cluster, as done in previous works (see \citealt{Billi+23,Billi+24}). In this case we do not find a clear relation between these two parameters. Fast and slow rotating BSSs do not have two distinct radial distributions.
\\
\citet{Ferraro+23a} found an intriguing relation between fraction of FR-BSSs and the host cluster central concentration and density, showing that low-concentration and low-density systems are populated by a larger fraction of fast spinning BSSs (see gray symbols in Figure \ref{fig:grafico_percentuale}). It remained unclear, however, whether such a dependence was either a sort of step function (with $\sim 40\%$ FR-BSSs for concentration parameters below $\sim 2$ and $\sim 10\%$ FR-BSSs above this limit), or a monotonically decreasing trend. The findings presented in this paper (blue circles in the figure) are an important addition to that result, indicating a continuous drop of the fraction of fast spinning BSSs for increasing $c$ and $\rho_0$. 

\begin{figure}
\centering
\includegraphics[width=\columnwidth]{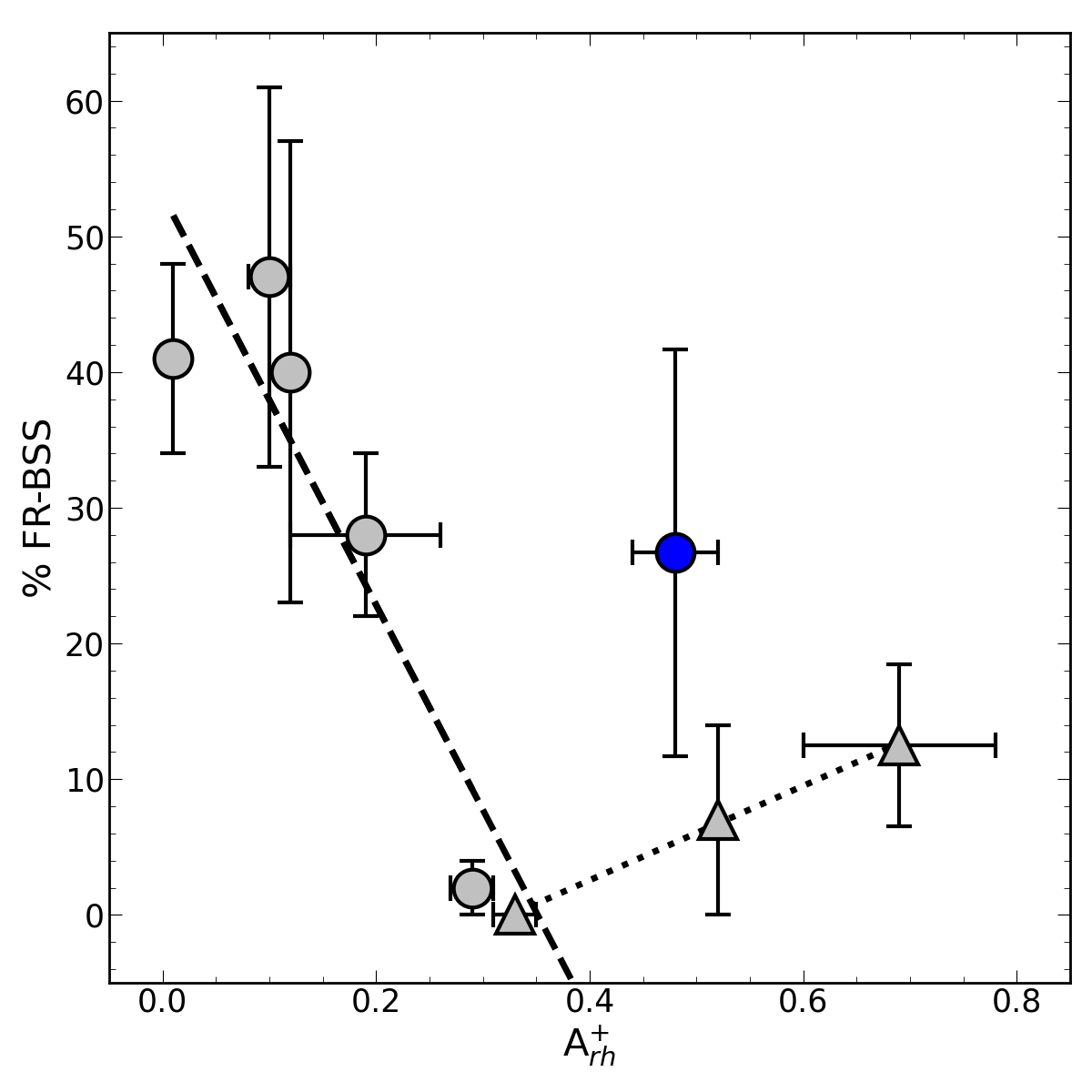}
\caption{Percentage of FR-BSSs as a function of the A$^+_{rh}$ parameter, which is a tracer of internal dynamical evolution of the host cluster. The gray symbols correspond to the GCs investigated in \citet{Ferraro+23a}, while the blue circle corresponds to NGC 1851.}
\label{fig:grafico_percentuale_a+}
\end{figure}

\noindent \citet{Ferraro+23a} also discussed a possible relation between the fraction of FR-BSSs and the A$^+_{rh}$ parameter, which is defined (\citealt{Alessandrini+16}; \citealt{Lanzoni+16}) as the area between the cumulative radial distribution of BSSs and that of a reference population of lighter stars (such as HB or RGB stars) and measures the dynamical age of the host cluster (see \citealp{Ferraro+18, Ferraro+23b}, and references therein): the larger is the value of A$^+_{rh}$, the higher is the cluster dynamical age, with A$^+_{rh}>0.3$ possibly indicating stellar systems that already suffered core collapse. The relation between the percentage of FR-BSSs and A$^+_{rh}$ found in \citet{Ferraro+23a} is plotted in with gray symbols in Figure \ref{fig:grafico_percentuale_a+} and it shows a clearly decreasing trend up to A$^+_{rh}\sim 0.3$, and a possible mild increase for larger A$^+_{rh}$ values. The latter behavior might testify the presence of recently formed COL-BSSs, generated by the strong density enhancement (hence, the growth in the stellar collision probability) occurred during core collapse. The result obtained in this work for NGC 1851 is superposed as a blue circle. The large error bar leaves open the possibility that the fraction of FR-BSSs increases as a function of A$^+_{rh}$ in highly dynamically evolved systems. However, in spite of its large value of the A$^+_{rh}$ parameter (A$^+_{rh}$ = 0.49, \citealt{Ferraro+18}), NGC 1851 is not classified as a post core collapse system. We thus speculate that the observed large percentage of FR-BSSs might testify recent BSS formation due to a large rate of binary interactions, which are also delaying the core collapse phase. Indeed, another possible clue of high rate of binary interactions in NGC 1851 is suggested by the presence of a relevant number of isolated millisecond pulsars and binaries that were likely formed via exchange interactions (\citealt{Ridolfi+22}, \citealt{Barr+24}, \citealt{Dutta+25}). 
\\
Unfortunately, these conclusions are forcedly quite speculative and need for further observational studies of large sample of BSSs in clusters with different characteristics. These require, however, high resolution and high S/N spectroscopy of relatively faint resolved stars in highly crowded field, thus representing a challenging task with the currently available instrumentation. Nevertheless the observed trends, especially those between the percentage of FR-BSSs and the central concentration and density of the parent cluster, call for a renewed theoretical effort aimed at providing a physical explanation for these observational results and shedding new light on the role of environment and internal dynamics in setting the BSS properties.

\begin{acknowledgements}
This work is part of the project {\it Cosmic-Lab} (Globular Clusters
as Cosmic Laboratories) at the Physics and Astronomy Department
``A. Righi'' of the Bologna University
(http://www.cosmic-lab.eu/Cosmic-Lab/Home.html). A.B. acknowledges
funding from the European Union - NextGeneration EU, Mission 4, Component 1, Investment 4.1 (D.M.
351/2022), CUP J33C22001300002. L.M. gratefully acknowledges support from ANID-FONDECYT Regular Project n. 1251809. A.M. acknowledges support from the project "LEGO – Reconstructing the building blocks of the Galaxy by chemical tagging" (P.I. A. Mucciarelli) granted by the Italian MUR through contract PRIN 2022LLP8TK\_001.
 
\end{acknowledgements}

\bibliographystyle{aasjournal}

\end{document}